\newcommand{\vdate}{March 1995}
\newcommand{\cernnr}{95-77}
\newcommand{\beq}{\begin{equation}}
\newcommand{\eeq}{\end{equation}}
\newcommand{\beqn}{\begin{eqnarray}}
\newcommand{\eeqn}{\end{eqnarray}}

\newcommand{\GeV}{\mbox{{ }GeV}}
\newcommand{\TeV}{\mbox{{ }TeV}}

\newcommand{\ZZ}{$\mbox{Z}^0$}
\newcommand{\WPM}{$\mbox{W}^\pm$}

\documentstyle[12pt,epsfig]{article}
\newlength{\sectionnumbersize}
\newlength{\sectionsize}
\begin{document}

\newcommand{\dgsa}[1]{\section{#1}}
\newcommand{\dgsb}[1]{\subsection{#1}}
\newcommand{\dgsm}[1]{\noindent{\Large\bf #1}}

\begin{titlepage}

\renewcommand{\thefootnote}{\fnsymbol{footnote}}
\setcounter{footnote}{0}

\hfill
\hspace*{\fill}
\begin{minipage}[t]{5cm}
\hfill CERN-TH/\cernnr
\end{minipage}

\vspace{0.5cm}
%\vspace{1cm}

\begin{center}

{\LARGE\bf
   {The Physics of the\\
    Standard Model Higgs Boson\\
    at the
    LHC}$\;${\it\footnote[2]
      {\em Invited talk presented
       at the XXXth Rencontres de Moriond, Les Arcs 1800,
       March 19-26, 1995; to appear in the proceedings of the conference
      }\\
   }
}

\vspace{1cm}
\vspace{0.5cm}

%\fbox{\fbox{D R A F T   1 . 1}}
%\vspace{0.5cm}

\vspace{0.1cm}

{\bf Dirk Graudenz}$\;${\it\footnote[1]{{\em Electronic
mail addresses: graudenz\char64{}cernvm.cern.ch,
i02gau\char64{}dsyibm.desy.de}} \\
\vspace{0.1cm}
Theoretical Physics Division, CERN\\
CH--1211 Geneva 23\\
}

\end{center}

\vspace{1.0cm}

\begin{abstract}
Some topics related to Standard Model Higgs boson physics at the Large Hadron
Collider are reviewed. Emphasis is put on an overview of QCD
corrections to Higgs boson decay and production processes.
\end{abstract}

\vfill
\noindent
\begin{minipage}[t]{5cm}
CERN-TH/\cernnr\\
\vdate
\end{minipage}
\vspace{1cm}
\end{titlepage}

\renewcommand{\thefootnote}{\arabic{footnote}}
\setcounter{footnote}{0}

\newpage
\dgsa{Introduction}
It is believed that the electroweak symmetry breaking
$\mbox{SU}(2)\times \mbox{U}(1)\rightarrow \mbox{U}(1)_{\mbox{em}}$
and the
generation of masses proceed via the
Higgs mechanism.
In the case of the electroweak Standard Model \cite{1},
the Higgs mechanism is
implemented by a weak isodoublet scalar field $\varphi$
with Lagrangian density (for a detailed review, see
\cite{2})
\beq
\label{higgsl}
   {\cal L}_{\mbox{Higgs}}=D\varphi^\dagger D\varphi
+\mu^2\varphi^\dagger\varphi-\lambda\left(\varphi^\dagger\varphi\right)^2.
\eeq
For $\mu^2>0$, the field $\varphi$ acquires a vacuum expectation
value $v=\sqrt{\mu^2/2\lambda}$. The potential of the
shifted field $\phi=\sqrt{2}(\varphi-v)$
has three flat directions and one direction
corresponding to a massive mode for small fluctuations around the minimum.
In the case of a global symmetry, the flat directions would correspond
to three massless Goldstone bosons. Owing to the local nature of the gauge
symmetry, however, these degrees of freedom
become longitudinal components of the gauge fields, which thereby
acquire a mass. The remaining massive mode is the physical Higgs boson.

The coupling of the Higgs boson to the electroweak gauge bosons
\ZZ{} and \WPM{} is due to the minimal coupling via covariant derivatives.
Fermions couple to the Higgs boson via
Yukawa couplings, where the coupling is strongest to heavy particles
like the $\tau$~lepton and the top and bottom quarks.

The vacuum expectation value of $\varphi$ is known
to be $v=246\GeV$. Limits on the mass of the Higgs boson can be inferred
from a possible triviality of the Higgs sector and from the requirement
of vacuum stability up to a scale $\Lambda$ \cite{3,4}.
For a top quark mass of $175\GeV$ and
for $\Lambda\approx 1\TeV$, the approximate bounds are
$70\GeV<m_H<500\GeV$. For $\Lambda\approx M_{\mbox{Planck}}$,
the bounds $125\GeV<m_H<200\GeV$ are obtained.

In the next section, the two main search strategies for the Standard Model
Higgs boson at pp colliders are reviewed.
In Section~\ref{QCD} we give a brief overview of QCD corrections
to decay and production processes of the Higgs boson, which have been
calculated during the last few years.

\dgsa{The Search for the Higgs Boson}
The LEP experiments have established a lower bound
on the Higgs boson mass of $m_H\geq63.8\GeV$ \cite{5}.
A lower bound of $85$--$90\GeV$ can be achieved at LEP2.
Higgs particles with larger masses can be produced at hadron colliders
(see \cite{6}, and for a recent review \cite{7}).

\begin{figure}[htb]
\centerline{\epsfig
{figure=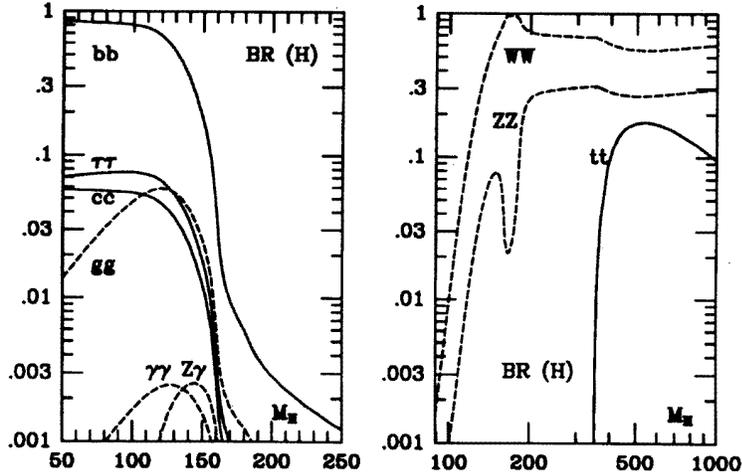,angle=0,width=10cm,clip=}
           }
{\em \caption{\label{decayrates} { Branching ratios of the Standard Model
Higgs boson}}}
\end{figure}

We begin by reviewing the branching ratios of the Higgs boson,
Fig.~\ref{decayrates}.
For $m_H>2m_Z$, the decay $\mbox{H}\rightarrow\mbox{ZZ}
\rightarrow\mbox{4l}$ provides a clean experimental signature.
For $130\GeV<m_H<2m_Z$, the decay
$\mbox{H}\rightarrow\mbox{ZZ}^*\rightarrow\mbox{4l}$ is still sizeable.
For $m_H<150\GeV$, the Higgs boson predominantly decays into
$\mbox{b}\overline{\mbox{b}}$-pairs. The experimental signature
for this process, however,
is swamped by QCD background. The only reliable signature is the
rare decay $\mbox{H}\rightarrow\gamma\gamma$, which is mediated by a
loop of charged heavy particles (\WPM-bosons and heavy quarks).

\begin{figure}[htb]
\centerline{\epsfig
{figure=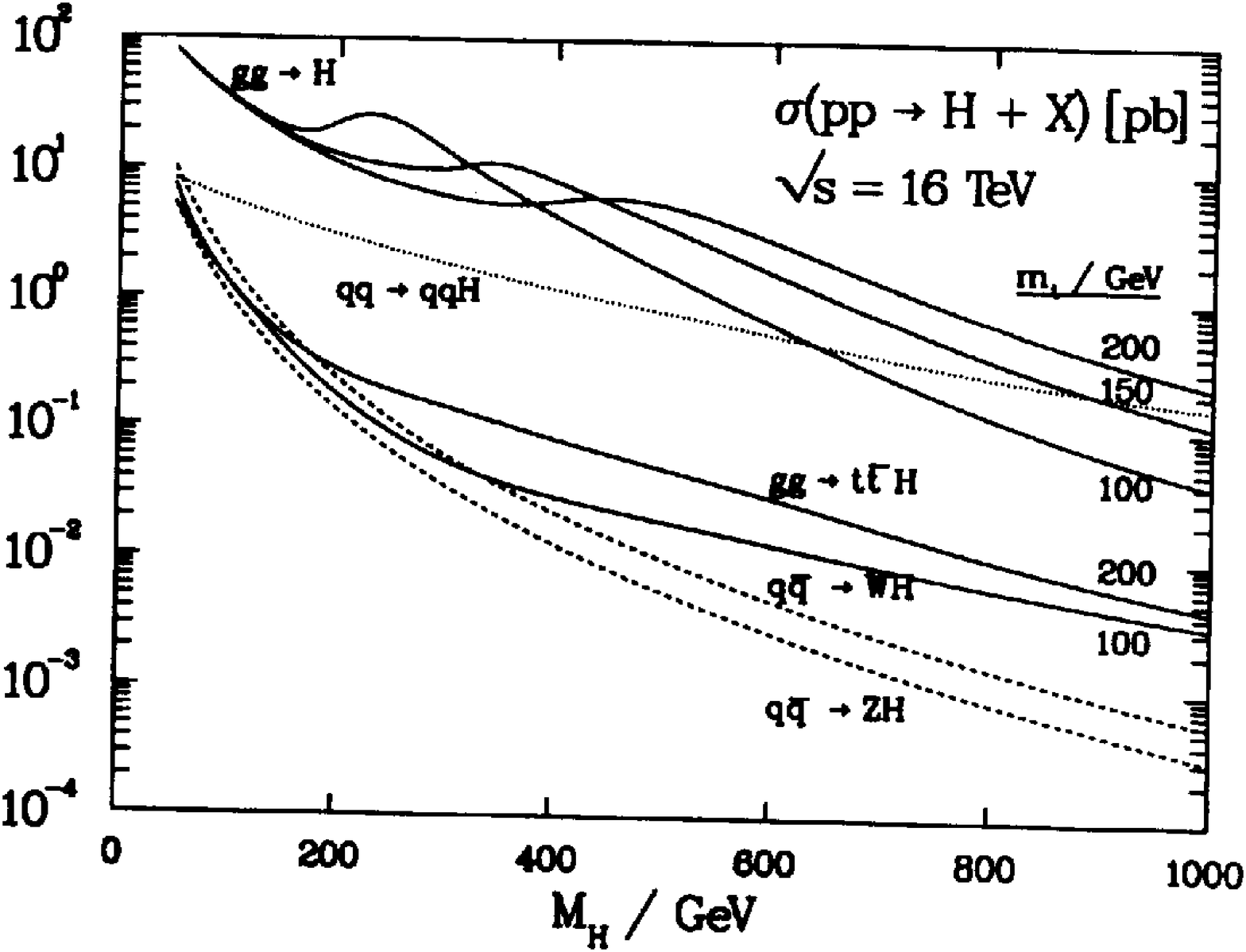,angle=1,width=10cm,clip=}
           }
\caption{\label{crosssection} Production cross sections
for the Standard Model Higgs boson in pp collisions }
\end{figure}

An overview of the production cross sections is given in
Fig.~\ref{crosssection} \cite{8}.
The main production mechanism of the Higgs boson at pp colliders
is the gg fusion mechanism, which proceeds via a heavy-quark triangle loop.
Other production mechanisms are $\mbox{WW}$ and $\mbox{ZZ}$ fusion,
where the $\mbox{W}$ and $\mbox{Z}$ bosons have been radiated from
the incoming quarks, $\mbox{t}\overline{\mbox{t}}$ fusion,
where two $\mbox{t}\overline{\mbox{t}}$ pairs are produced by incoming
gluons, and the radiation of a Higgs boson from a $\mbox{W}$ or $\mbox{Z}$
boson produced by $\mbox{q}\overline{\mbox{q}}$ fusion.

The main search strategies \cite{6} for the Higgs
boson at the LHC are, depending on its mass,
the production of the Higgs boson via gg fusion, and the observation of the
subsequent decays into $\mbox{ZZ}$ and $\mbox{ZZ}^*$, for the mass ranges
$m_H>2m_Z$ and $130\GeV<m_H<2m_Z$, respectively
(the experimental signature being four charged leptons from the
$\mbox{Z}$, $\mbox{Z}^*$ decays), and
the observation of
the decay into two photons, $\mbox{H}\rightarrow\gamma\gamma$,
for the mass range $90\GeV<m_H<130\GeV$.

The main background for the process
$\mbox{gg}\rightarrow\mbox{H}\rightarrow\mbox{ZZ}^{(*)}
\rightarrow\mbox{4l}$ are the processes
$\mbox{pp}\rightarrow\mbox{ZZ}^{(*)}+X
\rightarrow\mbox{4l}+X$ and
$\mbox{pp}\rightarrow\mbox{t}\overline{\mbox{t}}+X
\rightarrow\mbox{4l}+X^{\prime}$.
The latter can be reduced by requiring that $m_{l^+l^-}^2=m_Z^2$.
The process $\mbox{gg}\rightarrow\mbox{H}\rightarrow\gamma\gamma$
suffers from the large irreducible background
$\mbox{pp}\rightarrow\gamma\gamma+X$, which requires a very good
energy resolution for $\gamma$ pairs, and
from the reducible background $\mbox{pp}\rightarrow
\mbox{jet}+\mbox{jet},\,\mbox{jet}+\gamma$,
where the jets fake photons, which demands a detector with an
excellent photon/jet-discrimination capability.

More refined search strategies based on the observation of particles
or jets produced in association with the Higgs boson are discussed in
detail in \cite{9}.

\dgsa{QCD Corrections}
\label{QCD}
Here we briefly review the status of the calculation of QCD corrections
to the decay and production processes of the Standard Model Higgs boson.

The QCD corrections to the process $\mbox{H}\rightarrow\gamma\gamma$
decay width \cite{10} turn out to be small, of the order of
$1$--$2\%$ in the mass range $80\GeV<m_H<160\GeV$.
Results for the corrections to the decay
$\mbox{H}\rightarrow\mbox{gg}$ have been given in \cite{11} for the
case of an infinite loop-quark mass and in \cite{12} for the general
case. The corrections are of the order of $60$--$70\%$ for
$50\GeV<m_H<400\GeV$.
For the decay of the Higgs boson into heavy quarks, we refer
to the review in \cite{13}.

Now we discuss QCD corrections to the production processes.
The vector boson fusion process can be treated in a structure function
approach \cite{14}. The corrections are of the order of
$6$--$8\%$. The corrections to the process where the Higgs boson is radiated
from a $\mbox{W}$ or $\mbox{Z}$ boson can be inferred from the
corrections to the Drell--Yan process: they are of the order of $10\%$
\cite{15}.

\begin{figure}[htb]
\centerline{\epsfig
{figure=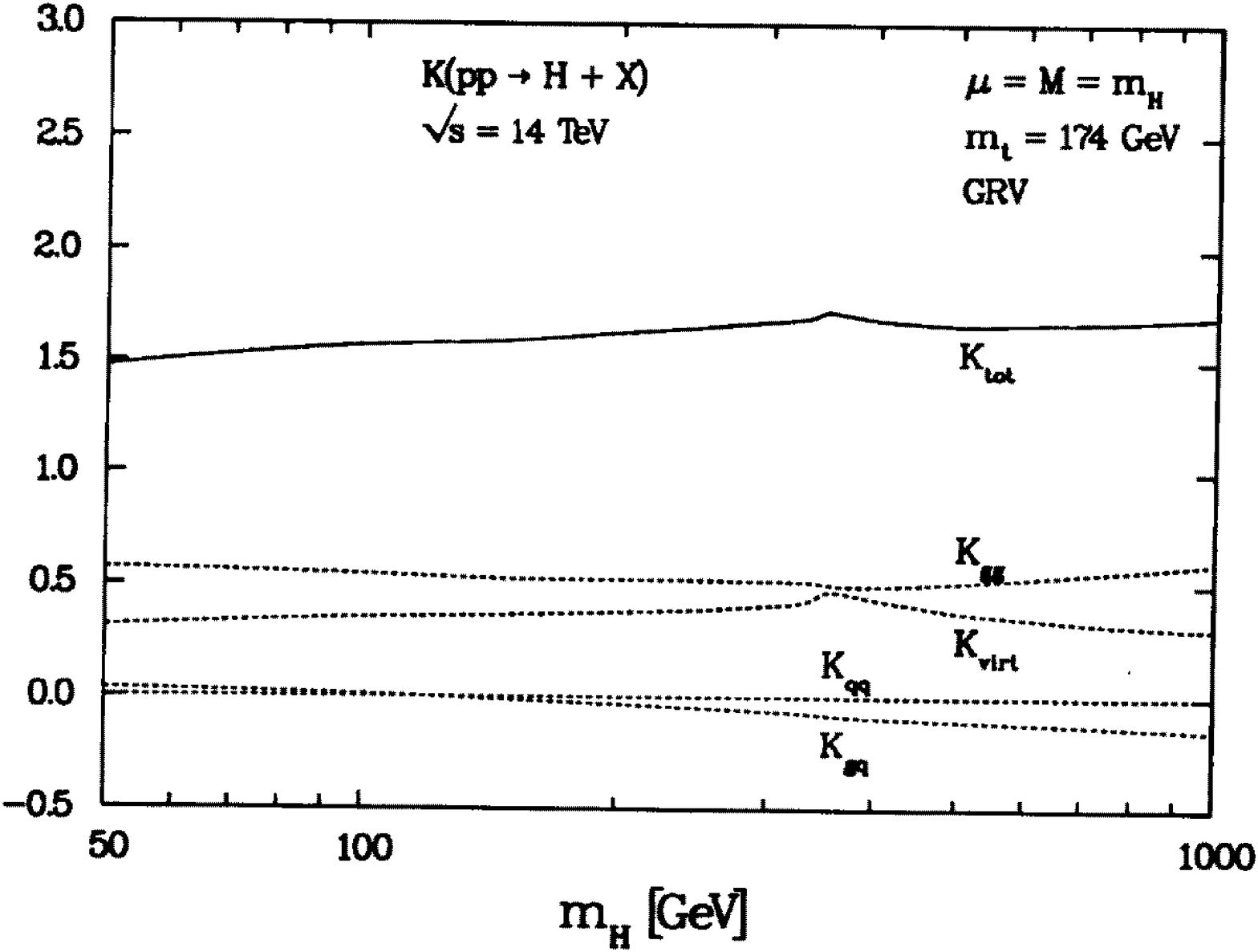,angle=1,width=10cm,clip=}
           }
{\em \caption{\label{kfactors} { K-factors for the process
$\mbox{gg}\rightarrow\mbox{H+X}$}}}
\end{figure}

We finally come to the corrections to the process $\mbox{gg}\rightarrow
\mbox{H}$. The first approaches have been in the limit for an infinite
loop-quark mass \cite{11,16} (see also \cite{17}).
The general case for arbitrary quark masses is treated in
\cite{18,12}.
The corrections are large, leading to a
K-factor\footnote{The K-factor is defined by
$\sigma_{\mbox{{\tiny NLO}}}
/\sigma_{\mbox{{\tiny LO}}}$,
where all quantities are consistently evaluated in leading and
next-to-leading order, respectively.}
of about $1.6$, see Fig.~\ref{kfactors}.
Thereby the dependence on the renormalization and factorization
scales is reduced by about $50\%$.

\dgsa{Summary and Conclusions}
The study of the
origin and nature of the electroweak symmetry-breaking mechanism
is one of the most important tasks to be performed at
the Large Hadron Collider.
We have given a very brief overview of the physics of the Standard
Model Higgs boson at
this machine.

If $m_H>2m_Z$, an elementary Higgs boson will certainly be found
via the process $\mbox{gg}\rightarrow\mbox{H}\rightarrow\mbox{ZZ}
\rightarrow\mbox{4l}$. For $130\GeV<m_H<2m_Z$, the search is more
difficult, and proceeds via $\mbox{gg}\rightarrow\mbox{H}\rightarrow\mbox{Z}
\mbox{Z}^*
\rightarrow\mbox{4l}$.
In the mass range $90\GeV<m_H<130\GeV$ the search is very difficult.
Here the rare decay $\mbox{H}\rightarrow\gamma\gamma$ has to be exploited.
In order to separate the signal from the background, a detector with
a very good energy resolution for $\gamma$ pairs and with an excellent
$\gamma$/jet-discrimination is required \cite{19}.

The QCD corrections to the main decay and production processes of the Standard
Model Higgs boson have been calculated during the last few years.
Because of their possibly large size and of the large scale dependence
of leading-order predictions, a solid study of the next-to-leading-order
corrections had been necessary, in particular for the processes
in the mass range where the search will turn out to be difficult.
By now, the QCD corrections are well understood and the scale dependence
is under proper control.

\vspace{5mm}
\dgsm{Acknowledgements}

\noindent
It is a great pleasure
to thank my collaborators M.~Spira and P.M.~Zerwas for valuable discussions.

% --- bibliography

%\newpage
\newcommand{\scs}{\rm}
\newcommand{\bibitema}[1]{\bibitem[#1]{#1}}
\newcommand{\bibbeginlong}{\begin{thebibliography}{XXX\,1988\,\,}}
\newcommand{\bibbeginshort}{\begin{thebibliography}{XX}}

\bibbeginshort %dummy

\addcontentsline{toc}{section}{References}

\bibitema{1}
      S.~Glashow, Nucl.\ Phys. \underline{20} (1961) 579;
      A.~Salam, in: {\it Elementary Particle Theory}, ed.\
      N.~Svartholm (1986); S.~Weinberg, Phys.\ Rev.\ Lett.\
      \underline{19} (1967) 1264.

\bibitema{2}
      J.F.~Gunion, H.E.~Haber, G.L.~Kane, S.~Dawson,
      {\it The Higgs Hunter's Guide},
      Addison Wesley, New York, 1990.

\bibitema{3}
      M.~Lindner, Z.~Phys. \underline{C31} (1986) 295.

\bibitema{4}
      G.~Altarelli, G.~Isidori,
      Phys.~Lett.\ \underline{B337} (1994) 141.

\bibitema{5}
      M.~Davier, in: Proceedings of the Lepton-Photon Conference,
      Geneva, 1991;
      J.~Schwid\-ling, in: Proceedings of the Int. Europhysics Conference,
      Marseilles, 1993, eds. J.~Car, M.~Perrotet.

\bibitema{6}
      Z.~Kunszt, W.J.~Stirling,
      in: Proceedings of the Large Hadron Collider ECFA Workshop,
      Aachen, 1990, eds. G.~Jarlskog, D.~Rein.

\bibitema{7}
      A.~Djouadi, Int.~J.~Mod.~Phys.~\underline{A10} (1995) 1.

\bibitema{8}
      M.~Spira, Ph.D.\ Thesis, Aachen, 1992.

\bibitema{9}
      L.~Poggioli, these proceedings.

\bibitema{10}
      A.~Djouadi, M.~Spira, J.~van der Bij, P.M.~Zerwas,
      Phys.\ Lett.\ \underline{B257} (1991) 187;
      H.~Zheng, D.~Wu, Phys.\ Rev.\ \underline{D42} (1990) 3760;
      S.~Dawson, R.P.~Kauffman, Phys.\ Rev.\ \underline{D47} (1993) 1264;
      K.~Melnikov, O.~Yakovlev, Phys.\ Lett.\ \underline{B312} (1993) 179;
      M.~Inoue et al., Mod.\ Phys.\ Lett.\ \underline{A9} (1994) 1189.

\bibitema{11}
      A.~Djouadi, M.~Spira, P.M.~Zerwas,
      Phys.\ Lett.\ \underline{B264} (1991) 440.

\bibitema{12}
      A.~Djouadi, D.~Graudenz, M.~Spira, P.M.~Zerwas,
      preprint DESY\ 94-123, CERN-TH/95-30,
      GPP-UdeM-TH-95-16 (1995).

\bibitema{13}
      L.~Surguladze, these proceedings.

\bibitema{14}
      T.~Han, G.~Valencia, S.~Willenbrock,
      Phys.\ Rev.\ Lett.\ \underline{69} (1992) 3274.

\bibitema{15}
      T.~Han, S.~Willenbrock,
      Phys.\ Lett.\ \underline{B273} (1991) 167.

\bibitema{16}
      S.~Dawson, Nucl.\ Phys.\ \underline{B359} (1991) 283.

\bibitema{17}
      S.~Dawson, R.P.~Kauffman,
      Phys.\ Rev.\ \underline{D49} (1993) 2298.

\bibitema{18}
      D.~Graudenz, M.~Spira, P.M.~Zerwas,
      Phys.\ Rev.\ Lett.\ \underline{70} (1993) 1372.

\bibitema{19}
      P.~Aurenche et al., in: Proceedings of the Large Hadron Collider
      ECFA Workshop,
      Aachen, 1990, eds. G.~Jarlskog, D.~Rein.

\end{thebibliography}

\end{document}